\documentclass[preprint2]{aastex}

\received{}
\revised{}
\accepted{}

\slugcomment{Version 1.3 : 20 December 2004}

\shorttitle{HIZSS3}
\shortauthors{Silva et al.}

\begin{document}

\title{The Distance and Metallicity of the Newly Discovered, Nearby
Irregular Galaxy HIZSS3\altaffilmark{1}}

\altaffiltext{1}{The optical spectroscopic observations reported here
were obtained at the MMT Observatory, a joint facility of the
Smithsonian Institution and the University of Arizona. The near-IR
imaging observations reported here were collected at the European
Southern Observatory, Cerro Paranal, Chile, within observing program
271.B-5047.}

\author{David R. Silva}
\affil{European Southern Observatory}
\affil{Karl-Schwarzschild-Str. 2, Garching bei M\"unchen, D-85748, Germany}
\email{dsilva@eso.org}

\and

\author{Philip Massey}
\affil{Lowell Observatory}
\affil{1400 W. Mars Hill Road, Flagstaff, AZ, 86001, USA}
\email{Phil.Massey@lowell.edu}

\and

\author{Kathleen DeGioia-Eastwood}
\affil{Department of Physics and Astronomy}
\affil{Northern Arizona University, P.O. Box 6010, Flagstaff, AZ, 86011-6010, USA}
\email{Kathy.Eastwood@nau.edu}

\and

\author{P. A. Henning}
\affil{Institute for Astrophysics}
\affil{University of New Mexico, 800 Yale Boulevard, NE, Albuquerque, NM, 87131-1156, USA}
\email{henning@as.unm.edu}

\begin{abstract}

HIZSS3 is an H~I source in the Zone of Avoidance. Its radio
characteristics are consistent with it being a previously unknown
nearby ($\sim$ 1.8 Mpc) low-mass dwarf irregular (dIm) galaxy. Optical
observations have shown that it contains a modest H~II region but
failed to reveal a resolved stellar population.  New spectroscopic
observations of the H~II region obtained at the MMT Observatory are
presented here.  They are used to derive the line-of-sight extinction
($E(B-V) = 1.41\pm0.04$) and gas metallicity (log $O/H + 12 \sim 7.8$)
of the H~II region. New near-IR imaging observations obtained at the
ESO Very Large Telescope are also presented here. These images clearly
reveal the resolved stellar population of HIZSS3 for the first
time. Narrow-band $P\beta$ images of the H~II region are used in
combination with previously published $H\alpha$ data to obtain an
independent line-of-sight extinction estimate: $E(B-V) =
1.32\pm0.04$. The adopted foreground extinction is $E(B-V) =
1.36\pm0.06$. Based on the K-band luminosity function and $K,J-K$
color-magnitude diagram, the apparent magnitude and color of the tip
of the red-giant branch (TRGB) are derived. In turn, these parameters
are combined with the adopted foreground extinction to estimate the
distance (1.69$\pm$0.07 Mpc) and mean red giant branch metallicity
($[Fe/H] = -0.5\pm0.1$). As an ensemble, these new observations
significantly strengthen the conclusion that HIZSS3 is a newly
discovered low-mass dIm lurking behind the Milky Way in the outskirts
of the Local Group.

\end{abstract}


\section{Introduction}
\label{sec:intro}

HIZSS3 ($l =$ 217.7, $b =$ 0.01) is an H I source in the Zone of
Avoidance (ZOA), i.e. the region near the Galactic plane where
foreground extinction from dust and gas causes an apparent decrease in
the surface density of extragalactic objects at optical wavelengths.
HIZSS3 was originally detected at 21-cm by the Dwingeloo Obscured
Galaxies Survey (Henning et al. 1998; Rivers, Henning, \&
Kraan-Korteweg 1999; Rivers 2000) and the H I Parkes Zone of Avoidance
Shallow Survey (Henning et al. 2000).  The name (HIZSS3) comes from
the list in the latter paper. Based on Very Large Array (VLA)
observations, it has an H~I mass of 2 $\times$ 10$^{7} {\mathrm
M}_\sun$, assuming a distance of 1.8 Mpc estimated from the
heliocentric velocity (280 km s$^{-1}$) of the H I source transformed
to the Local Group velocity centroid and assuming no peculiar velocity
(Massey, Henning, \& Kraan-Korteweg 2003, hereafter MHK2003). Given
the low radial velocity of the H~I gas, this distance (and hence mass)
estimate must be considered unreliable. A modest H$\alpha$ source at
the same heliocentric velocity and within the VLA 21-cm contours of
HIZSS3 was detected by MHK2003. This star-formation site is not
spatially coincident with the main, resolved H~I peak in the MHK2003
VLA 21-cm map, but rather with a semi-resolved secondary peak.  $BVRI$
images obtained by MHK2003 centered on the 21-cm peak failed to detect
definitively any stars associated with HIZSS3, likely due to a
combination of poor seeing ($\sim$ 3 arcsec) and high foreground
extinction (estimated A$_{\mathrm B} = $ 4.7 mags; Henning et
al. 2000).

As discussed by MHK2003, the characteristics of HIZSS3 suggest that it
is a newly discovered, nearby dwarf irregular galaxy (dIm).  At an
estimated distance of 1.8 Mpc, HIZSS3 lies beyond the zero-velocity
boundary of the Local Group.  However, this 21-cm redshift based
distance estimate relies on the assumption that HIZSS3 has negligible
peculiar velocity which is true for most, but not all (e.g. Leo I),
Local Group members. Thus, Local Group membership cannot be ruled out
based on existing radio data.

In this paper, new data are presented from the European Southern
Observatory Very Large Telescope and MMT Observatory facilities.  The
MMTO spectroscopic observations are used to measure foreground
reddening and the nebular metallicity of the H II region detected by
MHK2003. Foreground reddening is independently estimated using VLT
near-IR photometric observations and a different methodology.  The two
foreground reddening estimates are found to be essentially identical.
More significantly, the stellar component of HIZSS3 is directly
observed for the first time in near-IR images obtained with the VLT.
Using these images, near-IR luminosity functions as well as a
color-magnitude diagram (CMD) are constructed.  In combination, these
functions and the CMD are used to estimate the metallicity and
distance of the tip of the red-giant branch (TRGB).  In agreement with
MHK2003, the new derived properties of HIZSS3 presented here are
consistent with a previously unknown dwarf irregular galaxy on the
outskirts of the Local Group.

\section{Observations \& Data Processing}
\label{sec:obs}

\subsection{MMTO Observations}
\label{sec:obs-mmto}

Optical spectroscopy of the H~II region found in HIZSS3 by MHK2003 was
obtained on 2003 November 21 UT at the 6.5m MMT Observatory by PM and
KDE.  The primary goal of these observations was to determine
foreground reddening from spectrophotometry of H$\beta$ relative to
H$\alpha$, with a secondary goal of determining the oxygen abundance
if the [O II] $\lambda 3727$ nebular line could be detected despite
the high reddening.  The data were obtained with the MMT Blue Channel
spectrograph using a 500 l mm$^{-1}$ grating with a dispersion of
1.18\AA\ pixel$^{-1}$ using a 2688x512 backside-processed CCD provided
by the University of Arizona Imaging Technology Lab.  In order to
allow in more light, the slit was opened to 3.5 arcsec, resulting in a
resolution of 10 pixels (12\AA).  No blocking filter was used in order
to maximize the throughput, a seemingly safe choice given that
first-order H$\alpha$ would be contaminated by second-order light at
3280\AA.  While this proved to be a good choice for the H~II region,
the bluest flux standards (G191B2B and Feige 34) showed evidence of a
10\% contamination in the red compared to the flux standard Hiltner
600, which has a large Balmer jump.  Thus, only observations of the
latter were used for the flux calibration.  The HIZSS3 observations
were made as eight 1200 sec integrations, with the slit of the
spectrograph reset to the parallactic angle at the beginning of each
integration.

\begin{figure}
\epsscale{0.45}
\plotone{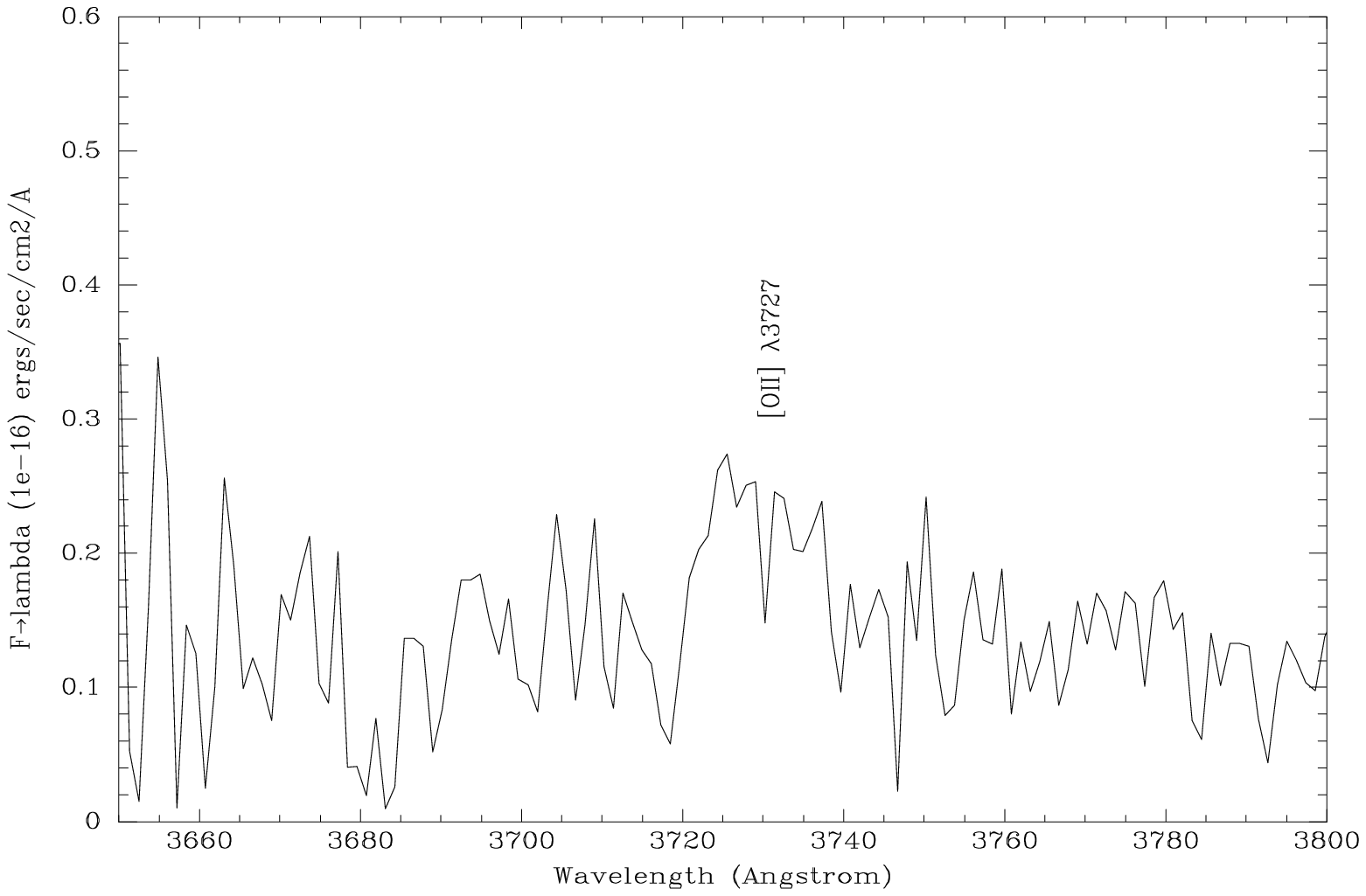}
\plotone{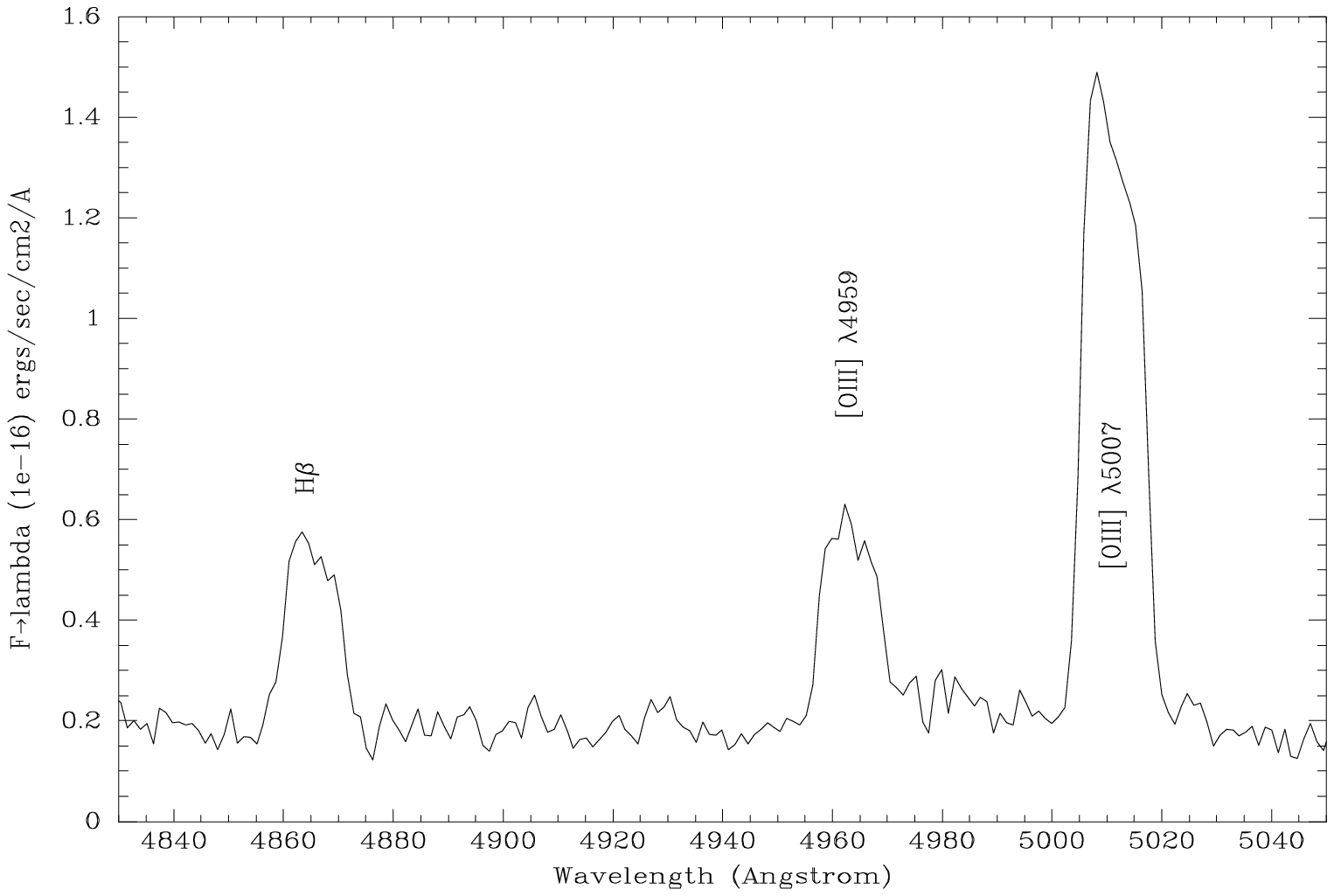}
\plotone{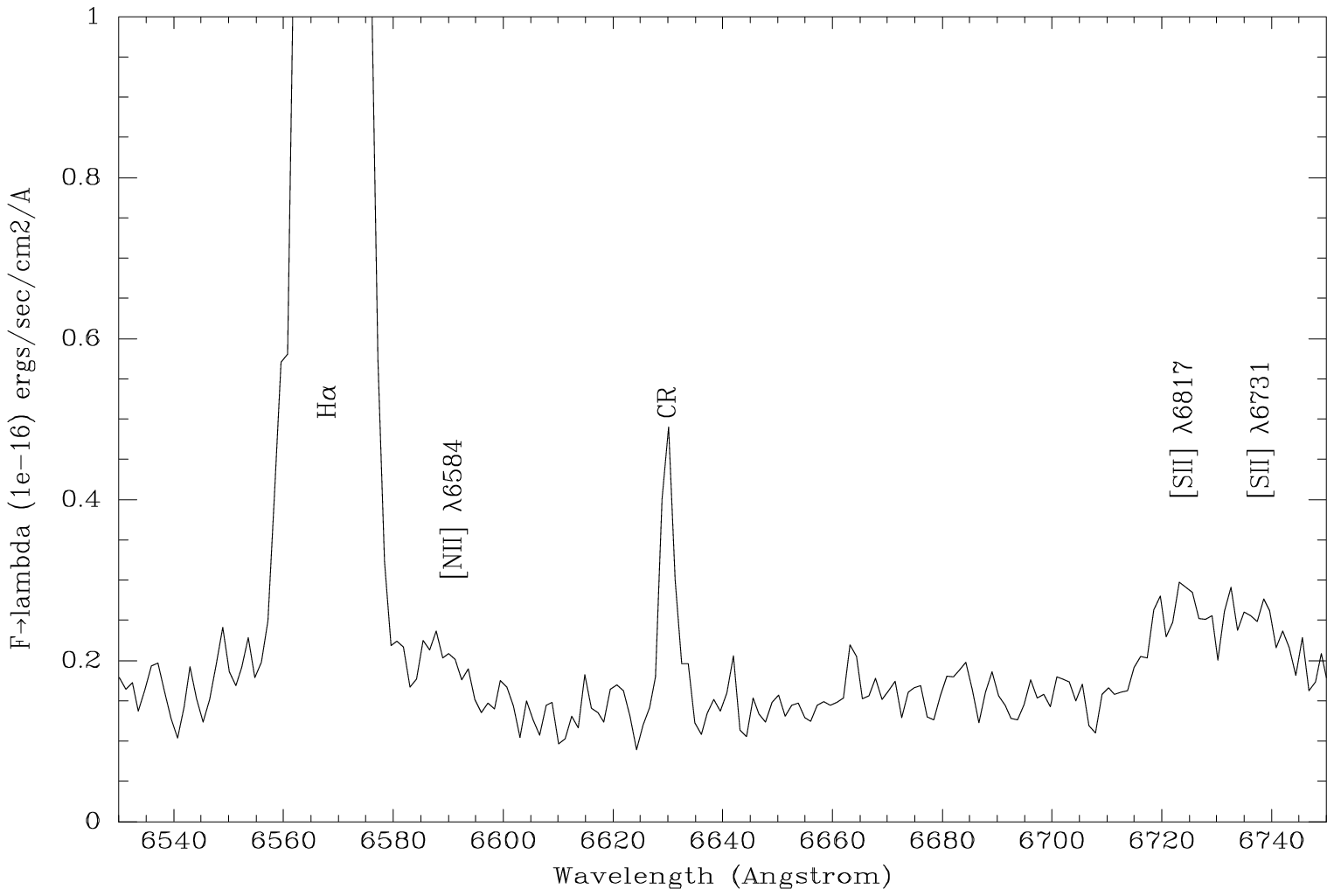}
\caption{\label{fig:spectra} Three spectral regions in our summed spectrum
of the H~II region in HIZSS3 are shown.
}
\end{figure}

IRAF\footnote{The Image Reduction and Analysis Facility (IRAF) is
distributed by the National Optical Astronomy Observatories, which are
operated by the Association of Universities for Research in Astronomy,
Inc., under contract with the National Science Foundation.}  was used
for all data processing.  Examination of the two-dimensional spectra
revealed that the spatial distribution of the nebular lines was
localized to 1 -- 2 arcsec, coincident with the H II region. The
spectra were extracted using the trace of an observation of the nearby
offset star as the template. Sky background was determined from each
side of the extraction aperture and subtracted. There is obvious
structure within the slit due to brightness variations of the nebula,
and this structure changed with the rotation angle.  Therefore, the
intensities of the strongest lines (H$\alpha$, H$\beta$, and [OIII]
$\lambda$ 5007) were measured from each individual spectrum.  The
spectra went deep enough to detect [OII] $\lambda 3727$, although this
line and the N[II] $\lambda 6584$ lines were so weak that they could
only be measured in the co-added spectrum. Three pertinent regions are
shown in Fig.~\ref{fig:spectra}

The relative emission-line strengths are given in
Table~\ref{tab:lines}.  The actual H$\alpha$ flux measured is only
about one-quarter the total H$\alpha$ luminosity measured from direct
images by MHK2003. This is due presumably to slit losses necessitated
by remaining at the parallactic angle (position angle $\sim 0$
degrees) that prevented the slit from being aligned with the H~II
region (position angle $\sim 130$ degrees; see Fig.~3 of MHK2003).

\subsection{VLT Observations}
\label{sec:obs-vlt}

\subsubsection{HIZSS3 and Reference Field}
\label{sec:vlt-images}

Near-IR imaging observations of HIZSS3 and a nearby reference field
were obtained at the ESO Very Large Telescope during 2003 November.
Time for these images was allocated within the ESO Director's
Discretionary Time (DDT) program. All VLT observations reported here
were executed in Service Mode.

Images were obtained using the short-wavelength (SW) arm of the ISAAC
near-IR imaging spectrometer mounted at one of the Nasymth foci of the
Antu/UT1 8.2m telescope. The SW arm uses a 1024x1024 Hawaii Rockwell
HgCdTe array with a plate scale of 0.148 arcsec pix$^{-1}$ for a
single-image field-of-view of approximately 2.5 $\times$ 2.5 arcmin.
ISAAC was operated in its standard jitter mode, where a series of
images is obtained with random telescope offsets of up to 15 arcsecs
relative to a fiducial center point between each individual
exposure.

A series of such jittered images was obtained for both the HIZSS3
field ($\alpha = 07^{\rm h}00^{\rm m}26^{\rm s}.9$, $\delta =
-04^{\circ}12'30''$, J2000) and a nearby reference field ($\alpha =
07^{\rm h}00^{\rm m}11^{\rm s}.0$, $\delta = -04^{\circ}03'38''$,
J2000) in the $J_s$ and $K_s$ filters. The HIZSS3 field center was
chosen to include both the 21-cm peak and the H~II region discussed in
MHK2003 (see that work for coordinates). The reference field was
selected to have the identical galactic latitude and contain no very
bright stars. It is otherwise unexceptional and located approximately
10 arcmins away.  An additional narrow-band $P\beta$ image (ISAAC
filter NB\_1.28) of the HIZSS3 field was obtained to allow the
determination of foreground reddening.  All images were obtained under
clear, apparently photometric, conditions (see photometric calibration
discussion below). A summary of these observations is provided in
Table~\ref{tab:vlt-obslog}.

The HIZSS3 and reference field image series were processed by the ESO
ISAAC pipeline into single images which were dark, bias, and
background corrected. These images were delivered as part of the
standard ESO Service Mode data package. Images were processed using
the {\it combine\_mc} algorithm, reported to produce photometrically
reliable final images\footnote{See
\url{http://www.eso.org/projects/aot/eclipse/jitterphot/}.}.
Photometric accuracy was verified by comparing the instrumental
magnitudes of stars detected in single frames with the same stars in
the combined frames. No systematic trends with respect to position or
magnitude were seen within the area where all images in the series
overlapped, confirming the photometric reliability of the pipeline
output.

Five output images were delivered: $J_s$ and $K_s$ images for both the
HIZSS3 and reference field pointings, as well as $P\beta$ for the
HIZSS3 pointing.  All five output images were trimmed to the same
area, corresponding to the output image with the smallest area where
all input images for that stack overlapped. The final size of the
trimmed images was 951 x 951 pixels, corresponding to 19,810
arcsec$^2$ or 5.503 arcmin$^2$.  The image quality of the output
images is excellent: the FWHM of a Gaussian fit to the FWHM is close
to 0.5 arcsecs in the $J_s$ and $K_s$ frames and 0.6 arcsecs in the
$P\beta$ frame.

These images clearly reveal the stellar content of HIZSS3 for the
first time. The $K_s$ HIZSS3 and reference field images are shown in
Figure~\ref{fig:Ks}.  The reference field image simply shows a uniform
distribution of stars, while the HIZSS3 field image clearly shows
enhanced object surface density at the locations of the H~II region
and the H~I peak.  Figure~\ref{fig:HII} shows zoomed $J_s$, $K_s$, and
$P\beta$ images of the H~II region. This latter figure clearly shows
stellar objects associated with the nebular gas.

\subsubsection{Photometric Calibration}
\label{sec:vlt-standards}

The ESO calibration program provides nightly observations of UKIRT
photometric standards (Hawarden et al. 2001) for ISAAC SW imaging
observations.  Additional observations of red UKIRT standards were
obtained in $J_s$ and $P\beta$ on the same nights that the HIZSS3
$P\beta$ observations were obtained. A summary of the relevant
standard stars observations is found in Table~\ref{tab:vlt-obslog}.

Input images of $J$ and $K$ UKIRT standards for all three nights were
processed by the ESO pipeline into standard images that were dark,
bias, and background corrected.  These output images were delivered as
part of the standard ESO Service Mode data package. The $P\beta$
standard images were not processed by the ESO pipeline because the
NB\_1.28 filter is not supported by the standard pipeline. Therefore,
input images were processed manually to correct for bias, dark, and
(minimal) background.

The instrumental magnitudes of the $J$ and $K$ UKIRT standards were
measured using standard IRAF aperture photometry tools. Apertures of
radii 15 pixels (2.22 arcsecs) were used, with sky annulus radii 20
pixels (2.96 arcsecs) and widths 10 pixels (1.48 arcsecs).  Each star
was observed as a sequence of five observations at different physical
locations on the array -- essentially, center, upper left, upper
right, lower right, and lower left.  For each star, the mean and
standard deviation was computed from the five individual observations.
In all cases, the standard deviation was approximately 0.03 mag.  This
uncertainty is typical for observations for photometric standards with
infrared arrays.  It does not arise from photon counting related
statistical reasons but from a combination of uncertainties from
flat-fielding and illumination corrections, light losses from
inter-pixel gaps, bad pixels, and intra-pixel sensitivity
variations.

Photometric transformation zeropoints ($\zeta_i$) were computed from
each standard star observation using the following equation:

$$ {\zeta}_i = m^{cat}_i - m^{ins}_i - {\kappa}_iX$$

\noindent where $i$ is the filter, $X$ is the airmass, $m^{cat}_i$ is
the catalog magnitude, $m^{ins}_i$ is the mean instrumental magnitude
(i.e. the mean of the five individual measurements), and ${\kappa}_i$
is the adopted extinction coefficient.  For $J_s$ and $K_s$, telluric
extinction coefficients of --0.07 and --0.05, respectively, were
chosen to minimise the scatter between individual zeropoint
measurements.  These values are consistent with the mean values stated
in the {\it ISAAC User's Manual}.  Not enough standard stars were
observed to determine reliable color terms independently.  However,
the ISAAC color terms are thought to be negligible (C. Lidman, private
communication).

Final nightly zeropoints were computed by computing the mean (and
standard deviation) of the individual observations.  For 2003 November
7 and 9 UT, the $J_s$ zeropoints are $24.71\pm0.025$ (5 stars) and
$24.77\pm0.004$ (2 stars).  For 2003 November 8 and 9 UT, the $K_s$
zeropoints are $24.16\pm0.003$ (2 stars) and $24.14\pm0.003$ (2
stars).  Combined with the adopted extinction coefficients, these
zeropoints transform instrumental magnitudes into $J$ and $K$ in the
UKIRT system (Hawarden et al. 2001; see that work for transformations
to other systems).

The formal uncertainties of the nightly zeropoints do not provide a
definitive constraint on the photometric quality of the relevant
nights because of the relatively small number of standard stars
observed (see Table~\ref{tab:vlt-obslog}).  What can be said is: (1)
these measured zeropoint values are consistent with the values
reported on the ISAAC quality control Web site\footnote{\url{See
http://www.eso.org/qc/.}} over a long time period, once aperture
differences are taken into account; (2) measurements separated by
hours on the same night produce zeropoints consistent to within a few
percent; and (3) variations in atmospheric transmission measured by
the Paranal Ambient Site Monitor (ASM) were less than 0.02 mag in the
R-band during the entire night for all three nights\footnote{See
\url{http://archive.eso.org/asm/ambient-server}.  As Start-Night, use
the UT Date minus one (1) day. For Site, select Paranal.}.  In
conclusion, the nights during which these VLT observations occurred
appear to be have been photometric at the 0.02 mag or better level.

\section{Analysis}
\label{sec:analysis}

\subsection{Foreground Extinction}
\label{sec:extinct}

Relative emission-line strengths measured in the H~II region are given
in Table~\ref{tab:lines}.  Using the Cardelli, Clayton, \& Mathis
(1989) reddening law (A$_{\rm H\beta}=3.609E(B-V)$ and A$_{\rm
H\alpha}=2.535 E(B-V)$) and an intrinsic ratio
H$\beta$/H$\alpha$=0.350, the derived foreground reddening is
$E(B-V)=1.41\pm0.04$, where the error represents the standard
deviation of the mean for the 8 individual ratios.  Using this
extinction, corrected flux ratios are given in Table~\ref{tab:lines}.
The reddening-corrected line ratio [OIII] $\lambda 5007$ to [OIII]
$\lambda 4959$ is 2.8, in good agreement with the 2.9 value expected
on theoretical grounds.

The VLT P$\beta$ images of the H~II region can be used to make a
similar but independent measurement of the foreground reddening,
namely by comparing the flux of the H~II region at P$\beta$
(1.28$\mu$m) to that at H$\alpha$.  This technique has been recently
exploited in a study of spiral galaxies by Jones, Elston, \& Hunter
(2002).  The only complication is the lack of spectrophotometric
standards in the near-IR, necessitating the use of model atmospheres
with JHK standards.  We are indebted to L. Jones for correspondence
and advice on this subject.

The UKIRT standards (Hawarden et al.\ 2001) FS~112 (G0) and FS~121
(K3) were observed through the ISAAC narrow band 1.28$\mu$m filter
(NB\_1.28) as well as the broadband $J_s$ and $K_s$ filters.  The
Kurucz (1992) Atlas 9 $\log g=4.5$, $T_{\rm eff}= 6000^\circ$K and
4750$^\circ$K models were adopted for these two stars. For each model,
the average fluxes within the $J_s$ and $K_s$ filters were computed.
For each star, a conversion factor between model flux (in physical
units) and observed flux (in ADUs per second) was computed using the
Hawarden et al. catalog magnitudes, under the assumption that $J_s =
0$ and $K_s = 0$ for Vega. The conversion factors for the two stars
agreed within 10\%, providing confidence in this procedure.

These conversion factors were then used to transform observed NB\_1.28
flux into absolute flux. In order to determine an emission-line flux
for a line-source, the width of the filter must also be taken into
account, i.e. $\int T_\lambda {\rm d}\lambda/T_{\rm 1.28\mu m}=194$,
where $T_\lambda$ is the transmission of the filter as a function of
wavelength\footnote{See Jacoby, Africano, and Quigley (1987), but
beware of an error in equation 3, in which $F_N(i)$ is used in place
of the correct quantity $F_{(i)}$}.  The P$\beta$ count rate was then
measured after aligning the J$_s$ image to P$\beta$, and subtracting
after matching the PSFs by convolution and scaling.  The derived flux
at P$\beta$ is $9.88\times10^{-15}$ ergs cm$^{-2}$ s$^{-1}$.  The
observed ratio (P$\beta$/H$\alpha=8.7\times10^{-2}$)$_{\rm obs}=0.45$
(where the flux at H$\alpha$ has been taken from MHK2003) can be
compared to the intrinsic ratio
(P$\beta$/H$\alpha=8.7\times10^{-2}$)$_{\rm int}=5.71\times 10^{-2}$
(Osterbrock 1974). Combined with the Cardelli et al.\ (1989) reddening
law, the foreground reddening was found to be
$E(B-V)=1.32\pm0.04$. This is in good agreement with the value found
from the MMTO optical spectroscopy (1.41$\pm$0.04).

The computed mean and standard deviation of these two foreground
reddening measurements is $E(B-V) = 1.36 \pm 0.06$. If 1.25 microns
and 2.2 microns are adopted as the effective wavelengths of the $J_s$
and $K_s$ bands, respectively, the Cardelli et al. reddening curve can
be used to compute $A(J-K)$: $A_{1.25} = A_J = 0.874 \times E(B-V) =
1.19\pm0.06$ mag and $A_{2.2} = A_K = 0.352 \times E(B-V) =
0.48\pm0.06$ mag. Therefore, $A_J - A_K = E(J-K) = 0.522 \times E(B-V)
= 0.71\pm0.06$ mag.

Formally, these foreground extinction measurements are only valid for
the line-of-sight to the H~II region. It is quite possible the
foreground extinction across the HIZSS3 field is variable, either due
to variations in the foreground Galactic dust distribution or due to
(hypothetical) dust within HIZSS3 itself. For example, the H~II region
could be more dusty than the rest of HIZSS3. It is impossible to
constrain the magnitude of possible spatially variable foreground
reddening without detailed spectrophotometry of the individual stars
in HIZSS3. Thus, a single-valued $E(B-V)$ is adopted for the rest of
this paper. As discussed later, this does not have a significant
impact on the main conclusions of this work.

\subsection{HII Region Metallicity}
\label{sec:h2-metal}

Without detection of the temperature-sensitive [OIII]$\lambda 4363$
line, quasi-empirical relations between other emission line ratios and
oxygen abundance (the so-called $R_{23}$ method) may be used to
approximate the metallicity of the H~II region (Pagel et al.\ 1979).
Three approaches are possible\footnote{We thank D. Hunter for her
guidance in this matter.}:

\noindent
{\bf (1) Classical $R_{\rm 23}$ method.}  The $R_{\rm 23}$ ratio
$$R_{\rm 23}=\frac{ {\rm [OIII]} \lambda 5007 + {\rm [OIII]}\lambda 4959 + 
{\rm [OII]}\lambda 3727}{{\rm H}\beta}=5.50$$
has been calibrated by Skillman (1989) for the case that the oxygen
abundance is much lower than solar.  Using his formula results in
a value $\log O/H+12=7.5$.

\noindent
{\bf (2) $R_{\rm 23}$ method modified by [OIII]/[NII].}  Edmunds \&
Pagel (1984) use a modified version of the $R_{\rm 23}$ method, which
relies upon using the strength of the N[II]$\lambda 6584$ line as an
additional calibration of $R_{\rm 23}$. The current measurements
result in
$$\log \frac{{\rm [OIII]}\lambda 5007 + {\rm [OIII]}\lambda
4959}{{\rm[NII]} \lambda 6584} =2.22,$$ yielding a value of $\log
O/H+12=7.8$.

\noindent
{\bf (3) $R_{\rm 23}$ method modified by [OIII]/[OII]}. McGaugh (1991)
instead calibrates the $R_{\rm 23}$ value with the relative strengths
of [OIII] and [OII]:
$$\log \frac{{\rm [OIII]}\lambda 5007 + {\rm [OIII]}\lambda 4959}{{\rm[OII]} 
\lambda 3727}=0.35.$$  This yields a value of  $\log O/H+12=7.85$.

In summary, $\log O/H+12\sim 7.8$, comparable to other metal-poor
irregular galaxies in the Local Group, i.e., IC 1613 ($\log$
O/H+12=7.85; Talent 1980), WLM ($\log$ O/H+12=7.77; Hodge \& Miller
1985), and Pegasus ($\log$ O/H+12=7.93; Skillman, Bomans, \&
Kobulnicky 1997). For comparison, the oxygen abundances of the LMC and
SMC are 8.37 and 8.13, respectively (Russell \& Dopita 1990), while
that of the solar neighborhood is 8.70 (Esteban \& Peimbert 1995).

\subsection{Photometric Properties of Resolved Stellar Population}
\label{sec:lumfunc}

\begin{figure}
\epsscale{1.1}
\plottwo{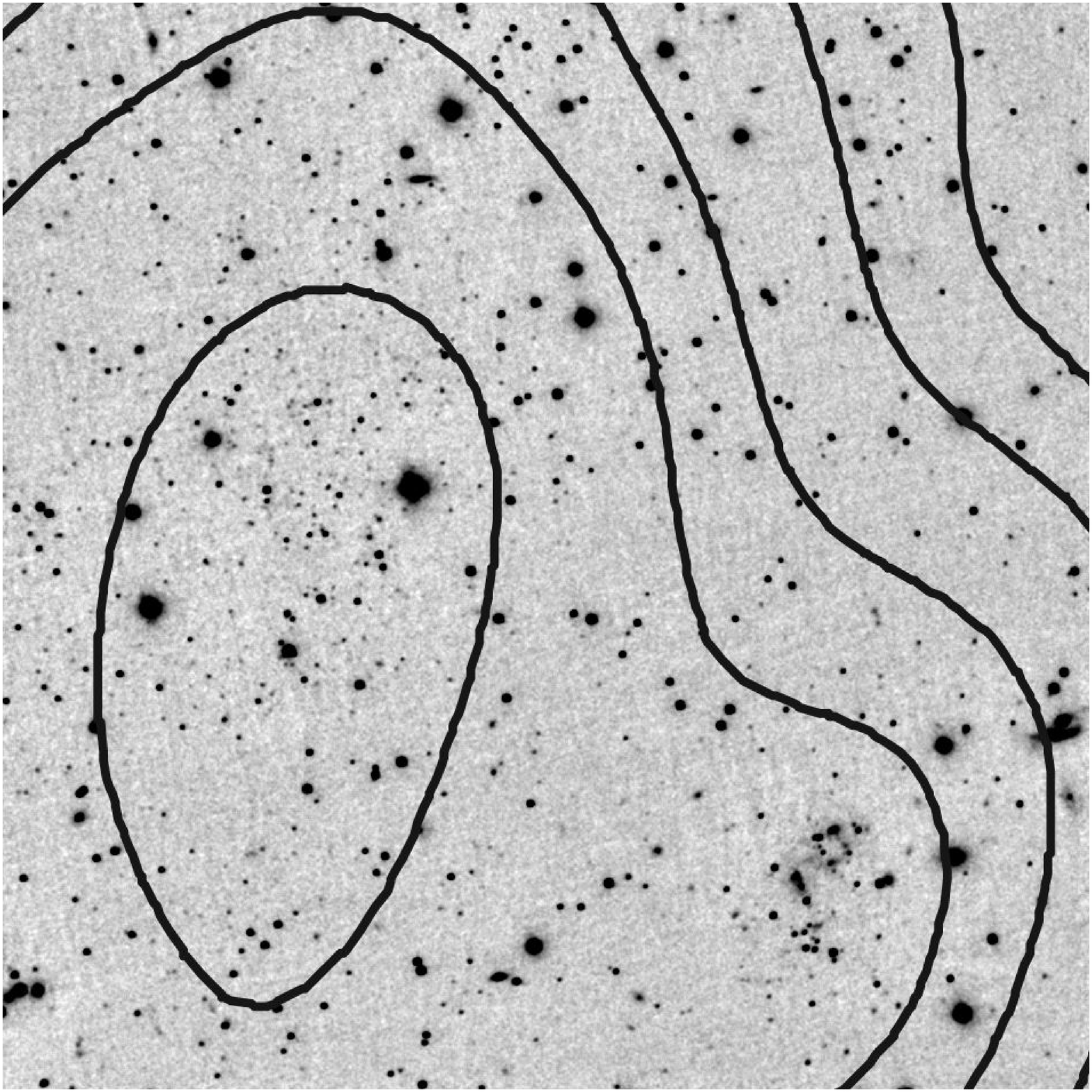}{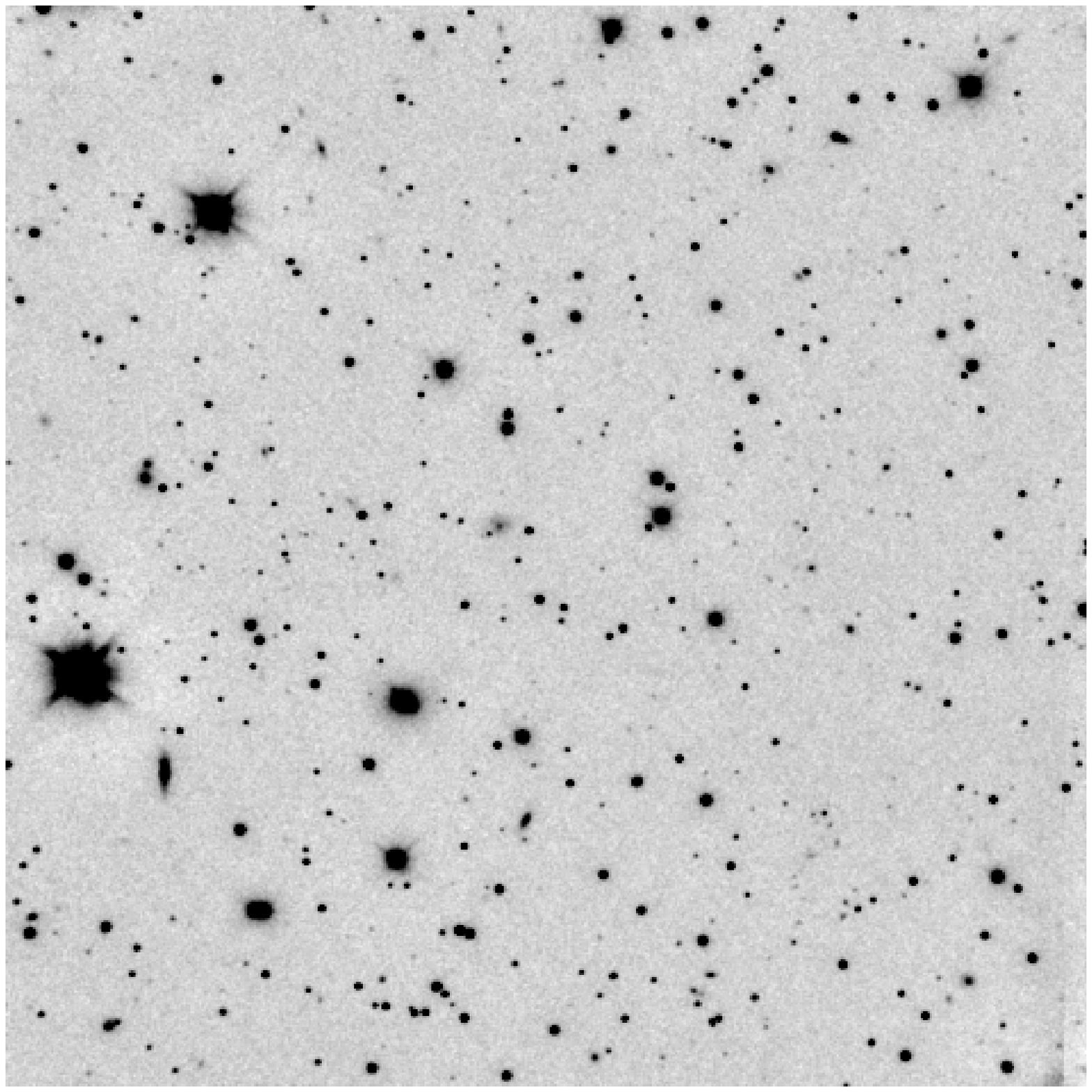}

\caption{\label{fig:Ks} $K_s$ HIZSS3 (left) and reference field
(right) images. North is up, east is left. Each image is 2.35x2.35
arcmin in size. In the HIZSS3 image, the H~I contours from MHK2003 are
overlaid (see their Figure 1). From inspection, it is clear that there
is a faint stellar population associated with the H~I peak while the
H~II region lies within a bulge in the H~I contours.  Objects in the
reference field are more uniformly distributed and less numerous.}
\end{figure}

\begin{figure}
\epsscale{0.3}
\plotone{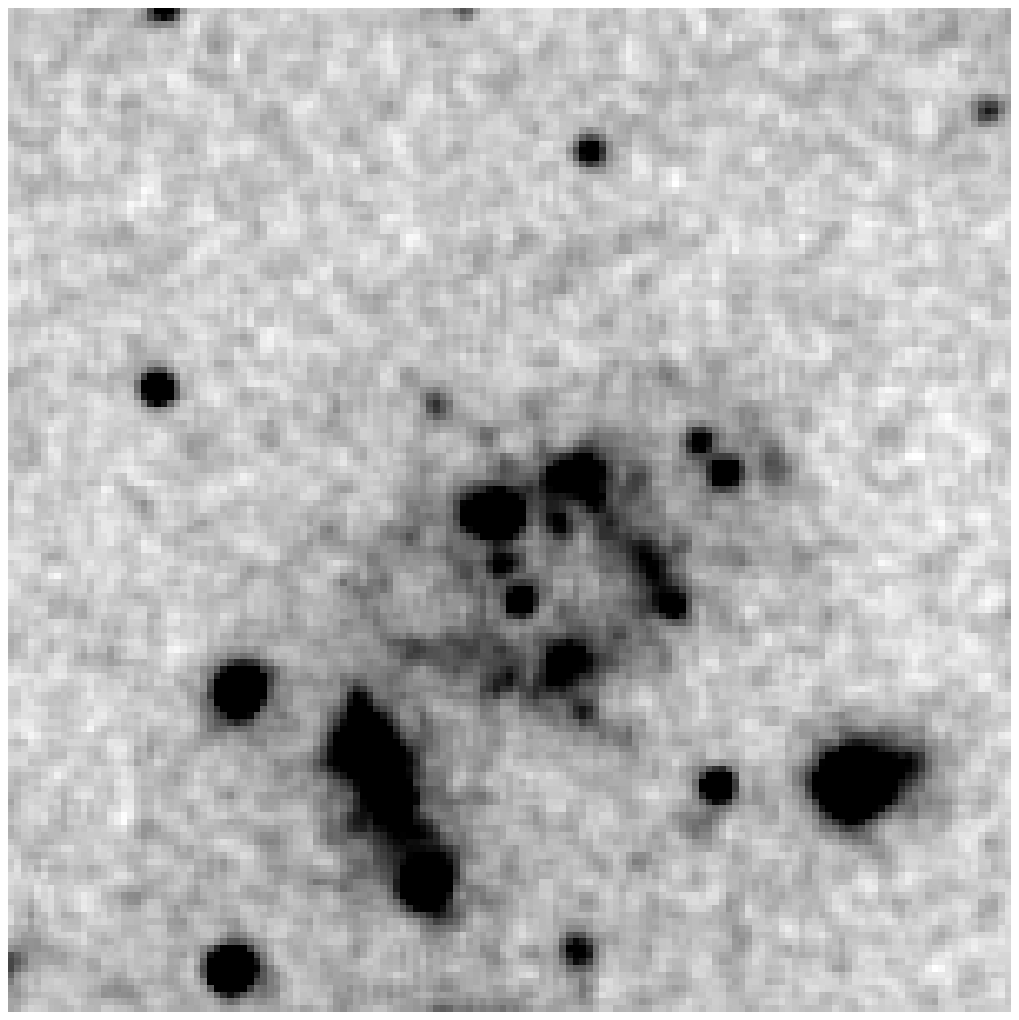}
\plotone{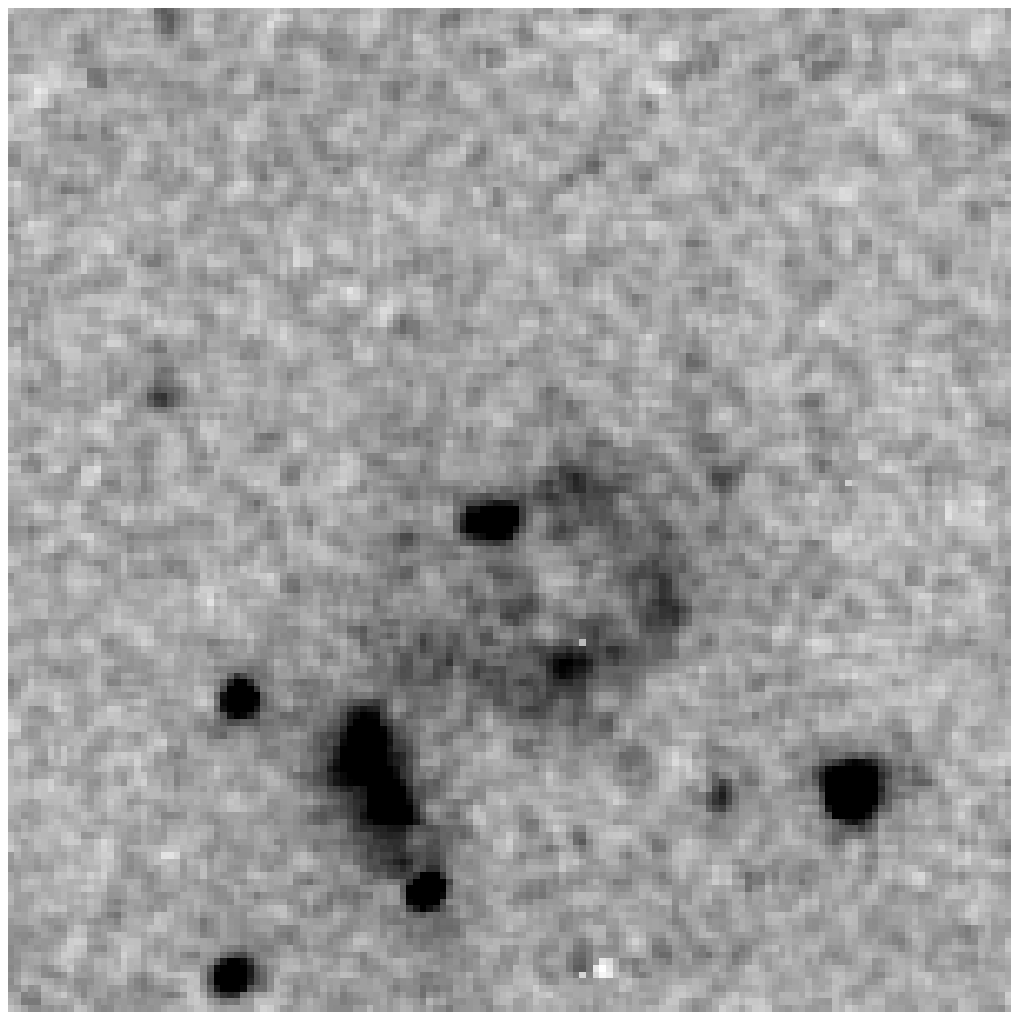}
\plotone{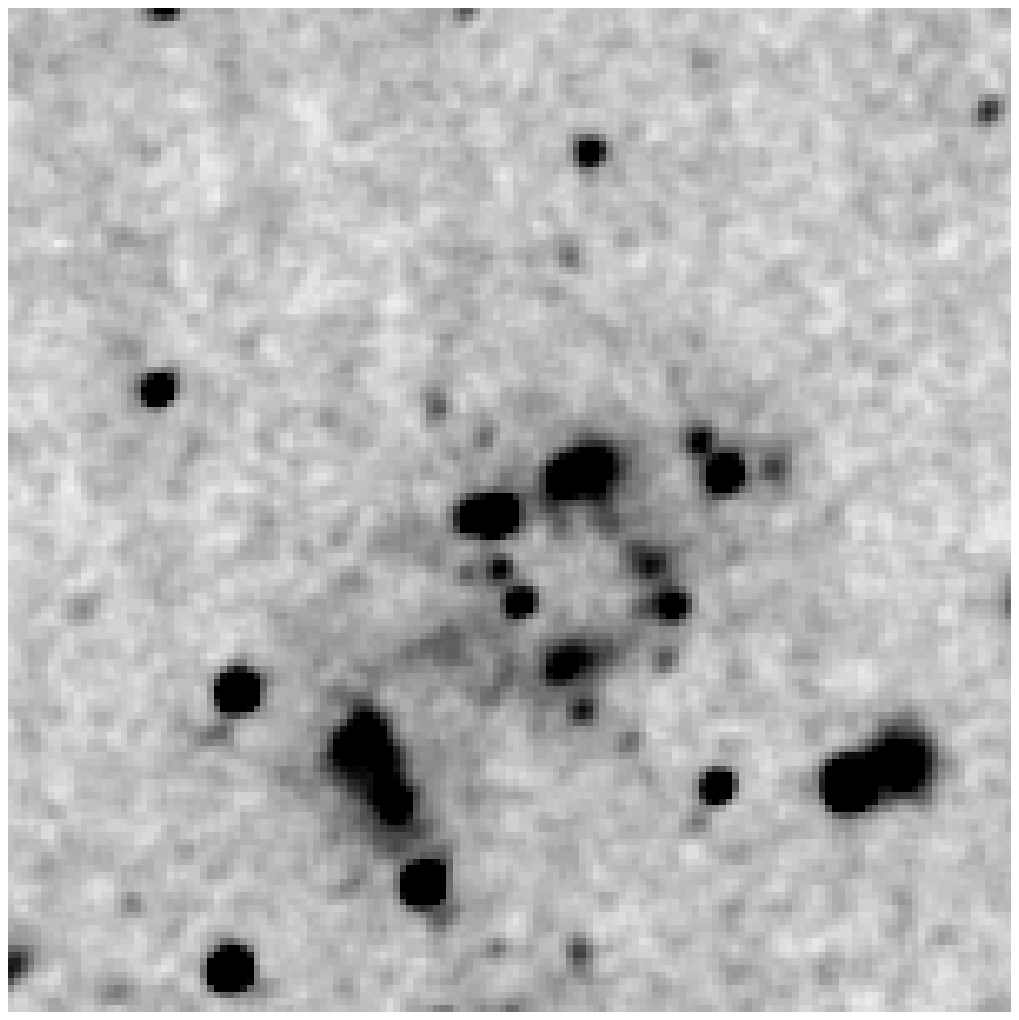}
\caption{\label{fig:HII} Cut-out images of the H~II region detected by
MHK2003. From left-to-right: $J_s$, $P\beta$, and $K_s$. North is up,
east is left. Each image is $22\times22$ arcsecs.}
\end{figure}

Stellar objects were identified in the $K_s$ HIZSS3 and reference
field images images using the DAOPHOT star finding routine implemented
under IRAF.  This routine identifies features that are significantly
above the local noise (in this case, 3$\sigma$) and whose roundness
and sharpness are typical of stars.  After removing objects that fell
partially on the edge of the images, 528 objects were detected in the
reference field and 739 in the HIZSS3 field.  Although a small
percentage of these stellar objects are undoubtably unresolved
background objects at high redshift, we assume henceforth that all the
detected objects are stars in the Milky Way or HIZSS3.  As can be seen
in Figure~\ref{fig:Ks}, the stellar surface density in the HIZSS3
field is higher near the positions of the H~I peak and H~II regions
reported by MHK2003.  In the reference field, the stellar surface
density is more uniform.

Instrumental aperture magnitudes of the detected stars were measured
using IRAF aperture photometry tools. Point spread function matching
was deemed unnecessary given the relatively low stellar surface
density (see Figure~\ref{fig:Ks}).  Instrumental magnitudes were
measured through a 5-pixel (0.74 arcsec) radius circular aperture,
using background measurements local to each star.  Using the same
aperture properties and the pixel coordinates of the stars detected on
the $K_s$ frames, aperture magnitudes were also measured on the $J_s$
frames after each $J_s$ frame was geometrically aligned with the
corresponding $K_s$ frame to sub-pixel accuracy.  A small number of
objects in the $K_s$ reference and HIZSS3 images (3 and 12,
respectively) were so faint in the corresponding $J_s$ images that
$J_s$ magnitudes could not be measured.

These instrumental magnitudes were transformed to the UKIRT system
using the photometric zeropoints and extinction terms described in the
last section.  An aperture correction from the 5-pixel radius used for
the program stars to the 15-pixel radius used for the standard stars
was computed using a few isolated stars in the program fields.  This
transformation process was repeated independently for all four HIZSS3
and reference field frames.

The result is a stellar catalog with $J$ and $K$ magnitudes (and
corresponding errors) in the UKIRT near-IR photometric system. Note
that these magnitudes have {\it not} been corrected for foreground
reddening.  In principle, such a correction is possible for stars
associated with HIZSS3 by ignoring possible variations in the
foreground dust distribution.  As shown below, however, stars can only
be assigned to HIZSS3 on a statistical basis, making {\it a priori}
extinction corrections of individual magnitudes impossible -- a
foreground extinction correction can only be made to ensemble
properties.  It is also impossible to correct foreground stars for
extinction because it cannot be assumed they lie at a common distance
and therefore a simple screen model cannot be assumed.

\begin{figure}
\epsscale{0.8}
\plotone{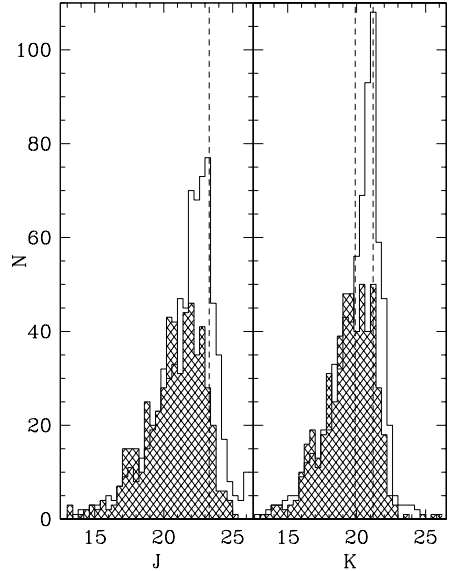}
\caption{\label{fig:lumFunc} $J$ and $K$ luminosity functions for the
stellar objects in the HIZSS3 and reference field images with $J$ and
$K$ magnitudes. The bin size is 0.4 mag. The hashed luminosity
functions contain the 525 stellar objects in the reference field. The
non-hashed luminosity functions contain the 727 stellar objects in the
HIZSS3 field.  The vertical dashed line in the $J$ panel indicates the
completeness limit in the HIZSS3 field.  The vertical dashed lines in
the $K$ panel indicate the adopted TRGB apparent magnitude and the
completeness limit in the HIZSS3 field.}
\end{figure}

Using this catalog, the luminosity functions shown in
Figure~\ref{fig:lumFunc} were constructed.  Bin widths are set to 0.4
mag in improve the statistical significance in each bin. In both
panels, the dashed (blue) lines are the 525 stars with $J$ and $K$
magnitudes in the reference field and the solid (black) lines are the
727 stars with $J$ and $K$ magnitudes in the HIZSS3 field.

\begin{figure}
\epsscale{1}
\plottwo{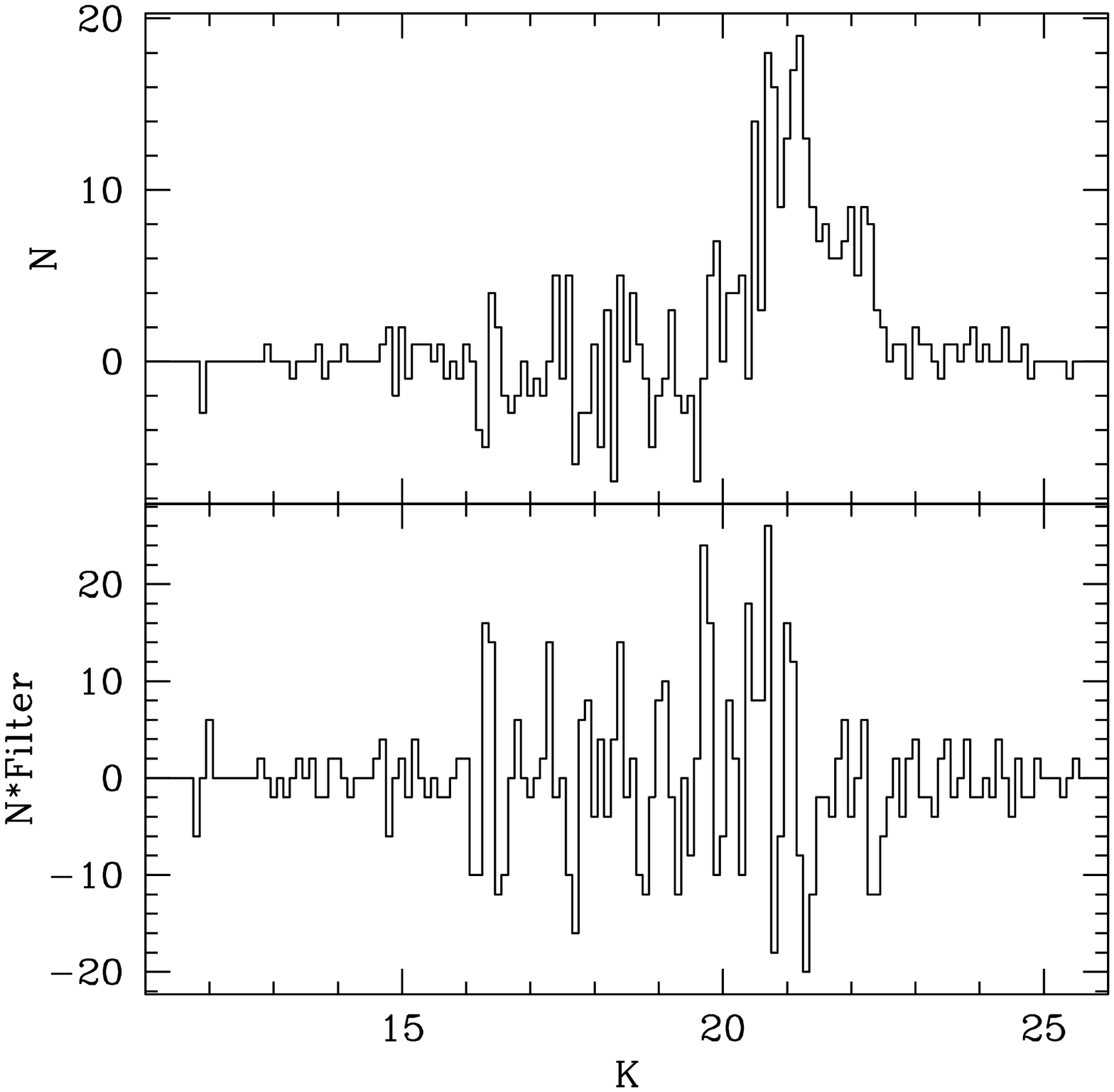}{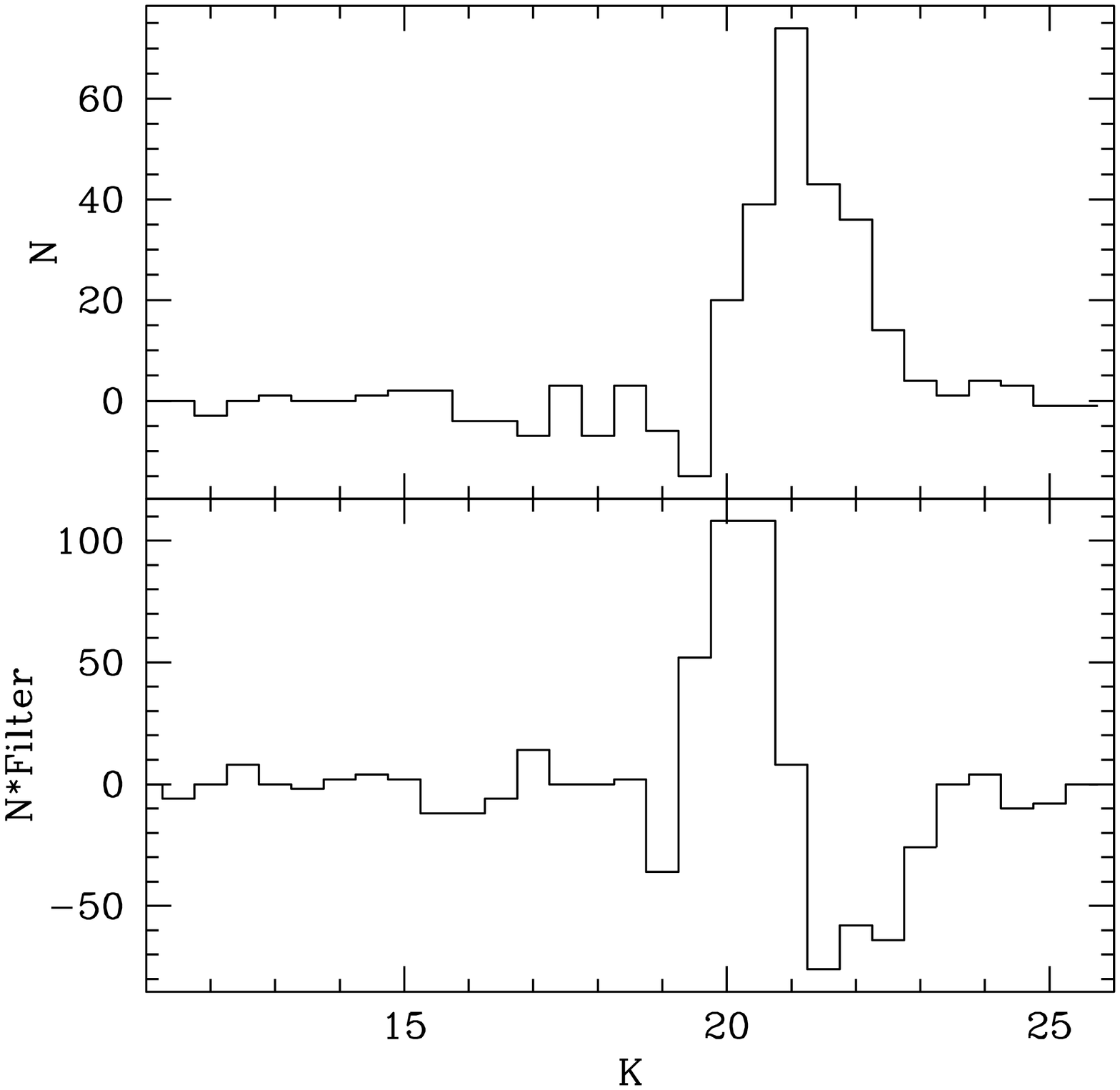}

\caption{\label{fig:edge} Background corrected K-band luminosity
function for HIZSS3 field before (top panels) and after (lower panels)
convolution with Sobel [$-$2,0,$+$2] edge detection filter.  Two bin
sizes are shown: 0.1 mag (left) and 0.5 mag (right). }

\end{figure}

Since HIZSS3 contains only one modest H~II region, star formation is
clearly not happening throughout the bulk of the galaxy.  Thus, the
brightest stars in the near-IR should be stars at or near the tip of
the first-ascent giant branch (TRGB). In principle, the apparent
magnitude and color of the TRGB can be used to constrain distance and
metallicity.  To detect the TRGB, the methodology of Lee, Freedman, \&
Madore (1993) is adopted. For a range of histogram bin sizes, the
HIZSS3 and reference field $K$ luminosity functions are constructed
and then subtracted to produce a background corrected $K$ luminosity
function for HIZSS3. Each background subtracted luminosity function is
then convolved with a 3-element edge detector filter with the values
[$-$2,0,$+$2] (i.e. the so-called zero-sum Sobel kernel).  The
resultant vector has local maxima at the luminosities where histogram
discontinuities occur.  The center of the {\it next} luminosity bin is
then adopted as the TRGB apparent magnitude. Figure~\ref{fig:edge}
provides illustrative examples for bin sizes of 0.1 and 0.5 mag,
corresponding to TRGB edges at $K$ = 19.80 and 20.00, respectively.

TRGB edges were determined for bin sizes 0.10, 0.15, 0.20, 0.25, 0.30,
0.35, 0.40, 0.45, and 0.50 mag.  For these nine bins, the mean and
standard deviation of the TRGB edge brightness is $K =
19.90\pm0.09$. This value is consistent with a visual inspection of
the luminosity functions shown in Figure~\ref{fig:lumFunc}. After
correction for foreground extinction ($A_{2.2} = 0.48\pm0.06$, see
Section~\ref{sec:extinct}), this value corresponds to $K_0 =
19.42\pm0.11$, where the edge brightness uncertainty and the
foreground reddening uncertainty have been added in quadrature.

From deeper near-IR observations in the literature, it is known that
object count should continue to increase with magnitude. This is
clearly not the case here, as shown in
Figure~\ref{fig:lumFunc}. Estimates of the apparent magnitude where
sample incompleteness becomes significant can be made by applying the
Sobel edge detection methodology again. For each field and each
filter, total luminosity functions (i.e. without background
subtraction) were constructed for bin sizes between 0.5 and 0.1 mag
using steps of 0.05 mag. Each luminosity function was then convolved
with the same 3-element edge detection filter as above.  For most
resultant vectors, significant incompleteness is indicated by the
global minimum. The {\it previous} (i.e. next brighter) bin is adopted
as the completeness limit. The mean and standard deviation over all
bin sizes are then computed for each field and filter.

The $K$ completeness limits for the HIZSS3 and reference images were
found to be 21.18$\pm$0.09 and 21.34$\pm0.21$, respectively - that is,
the same within the errors. The corresponding $J$ completeness limits
are 23.33$\pm$0.11 and 23.26$\pm$0.21 -- again, the same within the
errors. In both filters, the uncertainty of the reference field
completeness limit is larger due to the detection of fewer sources.

\begin{figure}
\epsscale{0.8}
\plotone{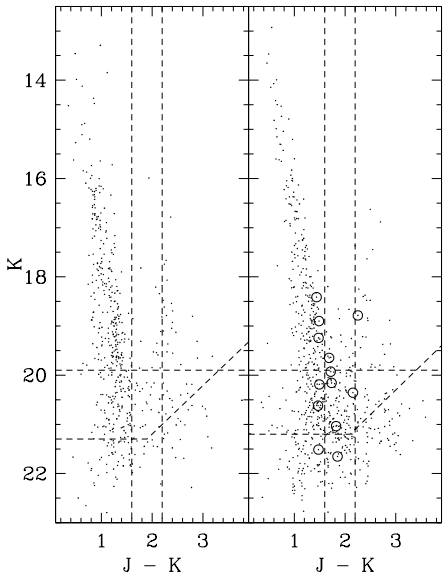}

\caption{\label{fig:cmd} Color-magnitude diagrams for reference (left)
and HIZSS3 (right) fields.  Only stars with $\sigma_{J-K} \leq 0.5$
mag are shown. The HIZSS3 TRGB edge at $K = 19.90\pm0.09$ is indicated
by a horizontal dashed line. The combination of the $J$ and $K$
completeness limits are shown as connected dashed horizontal and
diagonal lines. The vertical dashed lines indicate the color region
that contains the RGB candidates in the HIZSS3 field.  The circled
objects in the HIZSS3 (right) panel indicate stars that are spatially
coincident with the H~II region emission-line gas. See text for more
details.}

\end{figure}

The object catalog constructed above can also be used to construct the
HIZSS3 and reference field color-magnitude diagrams shown in
Figure~\ref{fig:cmd}. Only objects with $\sigma_{J-K} \leq 0.5$ are
shown.  In the HIZSS3 field, there are 44 objects with color
uncertainties larger than 0.5 mag, including 12 objects with no
measurable $J$ magnitude.  In the reference field, the corresponding
numbers are 8 and 3.

\begin{figure}
\epsscale{0.8}
\plotone{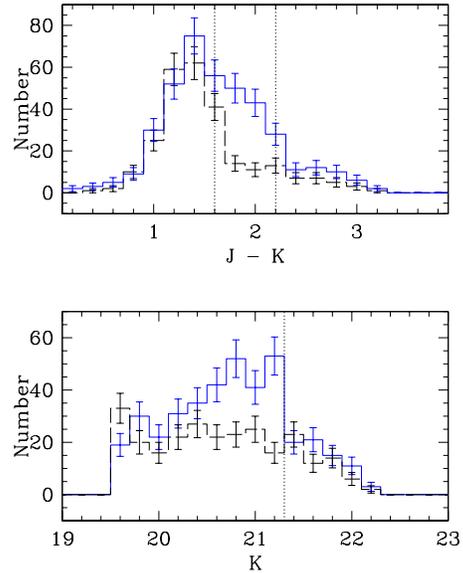}

\caption{\label{fig:colorHist} Color and luminosity distributions used
to select RGB candidates.  Only stars with $K > 19.5$ and
$\sigma_{J-K} \leq 0.2$ are shown. Neither color nor luminosity are
corrected for foreground extinction. In both panels, the solid (blue
in on-line version) histogram shows the HIZSS3 stellar distribution
while the dashed (black in on-line version) line shows the
distribution of objects in the reference field. In the color
distribution panel, the vertical dashed lines delineate the color
range of the candidate RGB stars in HIZSS3 (see text). In the
luminosity panel, the vertical dashed line shows the adopted K
completeness limit.}

\end{figure}

\begin{figure}
\epsscale{0.8}
\plotone{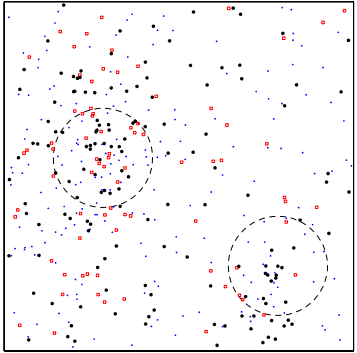}

\caption{\label{fig:xyPlot} Distribution of objects in HIZSS3 field
grouped by color. North is up, east is left, and the field-of-view is
2.35x2.35 arcmins in size (compare to HIZSS3 $K_s$ image shown in
Figure~\ref{fig:Ks}). Only objects with $K \geq 19.9$ and
$\sigma_{J-K} \leq 0.5$ are shown. Small filled (blue in on-line
version) circles correspond to $J-K < 1.6$. Large filled (black in
on-line version) circles correspond to $1.6 \leq J-K \leq 2.2$
(i.e. the HIZSS3 red giant candidates). Open (red in on-line version)
squares correspond to $J-K > 2.2$. The upper dashed circle is centered
on the VLA HI peak while the lower dashed circle is centered on the
HII region. Circle centers correspond to coordinates published by
MHK2003. The circle radii are 20 arcsecs. }

\end{figure}

Looking at Figure~\ref{fig:cmd}, it is clear that there are more stars
with $J-K \approx 1.9$ in the HIZSS3 field than in the reference
field. This is further quantified in Figure~\ref{fig:colorHist} which
shows that this excess lies in the approximate color range $1.6 < J-K
< 2.2$ for stars with $K > 19.5$ (i.e. slightly above TRGB and
fainter) and $\sigma_{JK} \leq 0.2$ mag. For these criteria, there are
significantly more objects (141) in the HIZSS3 field than in the
reference field (39).  These objects are adopted as red giant branch
(RGB) candidates in HIZSS3. The spatial distribution of these RGB
candidates is shown in Figure~\ref{fig:xyPlot}. The RGB candidates are
preferentially clumped near the VLA 21-cm HI peak and the HII region,
while bluer and redder stars are more uniformly distributed. The
median color of the RGB candidates in the HIZSS3 field is $J-K =
1.87\pm0.1$, corresponding to $(J-K)_0 = 1.16\pm0.12$ after extinction
correction where the color uncertainty has been added in quadrature
with the foreground extinction uncertainty. The 13 objects spatially
coincident with the H~II region emission-line gas are circled in
Figure~\ref{fig:cmd}. About half (6) of these objects are bluer than
the RGB candidates, consistent with them being hotter stars.

\subsection{Stellar Metallicity and Distance Estimates}
\label{sec:trgb}

Using the extinction corrected apparent magnitude of the TRGB ($K_0 =
19.42\pm0.11$) and adopted median TRGB color ($(J-K)_0 =
1.16\pm0.12$), the metallicity and distance to HIZSS3 can be estimated
by comparing the observed HIZSS3 properties to the properties of
Galactic globular clusters as presented by Valenti, Ferraro, \&
Origlia (2004a,b).  These papers are preferred over older papers by
the same team (e.g. Ferraro et al. 2000) because all the Galactic
globular cluster data have been transformed in a uniform way to the
2MASS photometric system.

Using the UKIRT-2MASS color transformations found on the 2MASS Web
site\footnote{\url{http://www.astro.caltech.edu/~jmc/2mass/v3/transformations/}},
the adopted median RGB color becomes $J-K_0 = 1.23$ on the 2MASS
system.  A similar UKIRT-2MASS transformation of $K_0$ is not
necessary given the small (0.003) color term in the published
transformation equation. The transformed median TRGB color is
consistent with the TRGB colors of the most metal-rich Galactic
globular clusters (see Valenti et al. 2004a, Figure 3) where $[Fe/H]$
lies in the range $-0.5\pm0.1$.  

Recall that in the H~II region, $[O/H]$ $\sim 7.8 - 8.7 = -0.9$,
i.e. the nebular gas appears to be significantly more metal-poor than
the stars near the TRGB. One speculative explanation for this
difference is that the star formation event traced by the H~II region
is being fueled by gas falling into HIZSS3 for the first time. In this
regard, it should be noted that the H~II region lies in a bulge of the
H~I gas (see MHK2003). It is possible that this bulge represents
recently arrived gas. Higher spatial resolution velocity maps of the
H~I gas would be very helpful for investigating this possibility. Such
images are currently being obtained and analyzed.

From Valenti et al. (2004b), $M_{Ks,2MASS}^{TRGB} \sim -6.7\pm0.1$ for
metal-rich Galactic globular clusters.  Using $K_0 = 19.42\pm0.11$ as
the foreground extinction-corrected TRGB apparent magnitude (see
above), a HIZSS3 distance modulus of $26.12\pm0.14$ or $1.69\pm0.07$
Mpc is implied.  Note that $M_{K}^{TRGB}$ varies slowly with TRGB
metallicity and thus color (see, e.g., Valenti et al. 2004b, Figure
5).  The fact that the adopted TRGB color/metallicity is somewhat
uncertain does not have a major impact on the adopted
$M_{K}^{TRGB}$. Hence, it does not have a major impact on the distance
modulus uncertainty. The TRGB-based distance estimate is remarkably
consistent with the 21-cm redshift estimate of 1.8 Mpc published by
MHK2003.

Values for the TRGB apparent magnitude and median RGB color have been
set from the aggregate properties of stars distributed spatially
across the entire HIZSS3 field. However, a measurement of $E(B-V)$ is
only available along one line-of-sight, i.e. to the H~II region
contained within HIZSS3. If this H~II region is more dusty than the
rest of HIZSS3, the foreground extinction correction for the TRGB
apparent magnitude would be smaller, increasing the estimated distance
to HIZSS3. However, the corrected median RGB color would also become
redder, reaching values in excess of median RGB colors of Galactic
globular clusters, which seems unlikely. Spatially variable foreground
extinction is also possible, but its effects (if present) would be
limited to increasing intrinsic $E(B-V)$ uncertainty and thus distance
measurement uncertainty. In summary, the adoption of a single-valued
$E(B-V)$ does not have a significant impact on the measurement of mean
stellar distance or metallicity.

\section{Conclusions}
\label{sec:conclude}

Using data obtained at the MMT Observatory and ESO Very Large
Telescope, the following new properties of HIZSS3 have been derived:

\begin{enumerate}

\item{} A resolved stellar population has been revealed for the first
time.  Its near-IR properties are consistent with a TRGB apparent
magnitude of $K_0 = 19.42\pm0.11$ and a median TRGB color of $(J-K)_0
= 1.22\pm0.12$.

\item{} The line-of-sight extinction to the H~II region was found to
be $E(B-V) = 1.41\pm0.04$ based on long-slit spectroscopic
observations and $E(B-V) = 1.32\pm0.04$ based on narrow-band $H\alpha$
and $P\beta$ observations. Given the various uncertainties, this
agreement is excellent. The adopted foreground extinction is $E(B-V) =
1.36\pm0.06$ for the near-IR photometric measurements.

\item{} Based on emission-line equivalent width measurements, the H~II
region metallicity was found to be log $O/H + 12 \sim 7.8$,
corresponding to $[O/H] \sim -0.9$.  In contrast, the mean RGB
metallicity was estimated to be $[Fe/H] = -0.5\pm0.1$ based on TRGB
color. In other words, stars on the RGB appear to be somewhat more
metal-rich than the nebular gas. This difference could be explained if
the star formation event traced by the H~II emission has been fed by
lower metallicity gas falling into HIZSS3 for the first time. Higher
spatial resolution maps of the H~I gas have been obtained and will be
analyzed to investigate this possibility.

\item{} Using the derived TRGB apparent magnitude of $K_0 =
  19.42\pm0.11$ and an adopted absolute magnitude of $-6.7\pm0.1$, the
  HIZSS3 distance modulus was found to be $26.12\pm0.14$,
  corresponding to a physical distance of $1.69\pm0.07$ Mpc. This
  estimate is in excellent agreement with distance estimate of $\sim$
  1.8 Mpc derived from 21-cm redshifts under the assumption of zero
  peculiar velocity with respect to the Local Group velocity centroid.

\end{enumerate}

These newly derived quantities significantly strengthen the conclusion
that HIZSS3 is a newly discovered dIm lurking at the edge of the Local
Group. As discussed by MHK2003, HIZSS3 appears to be the nearest dIm
discovered in the last 25 years.  However, the TRGB distance estimate
presented here ($\sim$ 1.7 Mpc) supports the MHK2003 conclusion that
HIZSS3 is {\it not} a member of the Local Group, having more in common
with galaxies at a similar distance as Sextans A, Sextans B, NGC 3109,
and Antlia.

\acknowledgements

We thank the ESO Director's Discretionary Time Committee (DDTC) for
approving the ESO observations.  We also thank Lowell Tacconi-Garman,
Emmanuel Jehin, and Wolfgang Hummel for their end-to-end assistance
with the VLT observations. We gratefully acknowledge several useful
conversations with Deidre Hunter and Lauren Jones. PM acknowledges
support from NSF grant AST0093060. Finally, we thank the anonymous
referee for helpful comments and suggestions.

\facility{VLT(ISAAC)}
\facility{MMTO(Blue Channel Spectrograph)}

\clearpage

%
%

%
%

%
%

%
%

%
%
%

%
%

%
%

%
%

\clearpage

%
%
\begin{deluxetable}{l c c}
\pagestyle{empty}
\tabletypesize{\scriptsize}
\tablewidth{0pc}
\tablenum{1}
\tablecolumns{3}
\tablecaption{\label{tab:lines}Relative Emission-line Strengths} 
\tablehead{
\colhead{Spectral Line}
&\colhead{Uncorrected Flux} 
&\colhead{Corrected Flux}
}
\startdata
~[OII]$\lambda 3727$  & 0.033 & 0.60 \\
~[OIII]$\lambda 4959$ & 0.10  & 0.35 \\
~[OIII]$\lambda 5007$ & 0.29  & 0.98 \\
~H$\beta$             & 0.087 & 0.35 \\
~H$\alpha$            & 1.000 & 1.00 \\
~[NII]$\lambda 6584$  & 0.0080 & 0.0080\\
\enddata
\end{deluxetable}

%
%
\begin{deluxetable}{c c c c c c}
\pagestyle{empty}
\tabletypesize{\scriptsize}
\tablewidth{0pc}
\tablenum{2}
\tablecolumns{6}
\tablecaption{\label{tab:vlt-obslog}VLT Observation Log} 
\tablehead{
\colhead{UT Date$+$Time}
&\colhead{Target} 
&\colhead{Integration}
&\colhead{Filter}
&\colhead{Airmass}
&\colhead{Seeing}
\\
\colhead{} 
&\colhead{}
&\colhead{(s)}
&\colhead{}
&\colhead{}
&\colhead{(arcsecs)}
}
\startdata
2003 Nov 06 23:40& UKIRT FS29 & 5$\times$14.2 & $J_s$ & 1.12 & \nodata \\
2003 Nov 07 00:12& UKIRT FS6 & 5$\times$14.2 & $J_s$ & 2.30 & \nodata \\
2003 Nov 07 04:55& UKIRT FS6 & 5$\times$14.2 & $J_s$ & 1.18 & \nodata \\
2003 Nov 07 07:00& UKIRT FS112 & 5$\times$20.0 & $P\beta$ & 1.09 & \nodata \\
2003 Nov 07 07:11& HIZSS3 & 11$\times$180 & $P\beta$ & 1.11 & 0.60 \\
2003 Nov 07 07:54& UKIRT FS121 & 5$\times$20.0 & $P\beta$ & 1.09 & \nodata \\
2003 Nov 07 08:00& UKIRT FS121 & 5$\times$20.0 & $J_s$ & 1.10 & \nodata \\
2003 Nov 07 08:08& HIZSS3 & 30$\times$60 & $J_s$ & 1.11 & 0.53 \\
\tableline
2003 Nov 07 23:52& UKIRT FS29 & 5$\times$14.2 & $K_s$ & 1.30 & \nodata \\
2003 Nov 08 07:43& HIZSS3 & 35$\times$60 & $K_s$ & 1.07 & 0.49 \\
2003 Nov 08 09:01& UKIRT FS14 & 5$\times$14.2 & $K_s$ & 1.10 & \nodata \\
2003 Nov 08 09:27& UKIRT FS10 & 5$\times$14.2 & $K_s$ & 2.28 & \nodata \\
\tableline
2003 Nov 09 03:50& UKIRT FS6 & 5$\times$14.2 & $J_s$ & 1.15 & \nodata \\
2003 Nov 09 04:03& UKIRT FS6 & 5$\times$14.2 & $K_s$ & 1.15 & \nodata \\
2003 Nov 09 06:11& Off-Field & 35$\times$60 & $K_s$ & 1.21 & 0.50 \\
2003 Nov 09 07:10& Off-Field & 30$\times$60 & $J_s$ & 1.10 & 0.48 \\
2003 Nov 09 08:57& UKIRT FS14 & 5$\times$14.2 & $J_s$ & 1.10 & \nodata \\
2003 Nov 09 09:09& UKIRT FS14 & 5$\times$14.2 & $K_s$ & 1.10 & \nodata \\

\enddata

\tablecomments{UT Time$+$Date indicates start of observing
sequence. Seeing is reported for final combined images produced by
ISAAC pipeline.  Seeing is not reported for standard stars.}

\end{deluxetable}

\end{document}